\documentclass{article}
\usepackage[latin1]{inputenc}
\usepackage[english]{babel}
\usepackage{dsfont}
\usepackage[cyr]{aeguill}
\usepackage{amsmath,amssymb}
\usepackage{amsthm}
\usepackage{eurosym}
\usepackage{multicol}
\usepackage{color}
\usepackage{fancyhdr}
\usepackage[pdftex]{graphicx}
\usepackage[pdftex]{hyperref}
\usepackage{verbatim}
\usepackage{setspace}
\usepackage{stmaryrd}

\newcommand{\indic}{\mathds{1}} 
\newcommand{\E}{\mathbb{E}} 
\newcommand{\sign}{\text{sign}} 

\theoremstyle{plain}

\providecommand{\keywords}[1]{\textbf{\textit{Keywords --}} #1}

\title{Efficiency of the financial markets \\ during the COVID-19 crisis: \\ time-varying parameters of fractional stable dynamics}

\author{Ayoub Ammy-Driss$^{\text{a}}$, Matthieu Garcin$^{\text{b,}}$\thanks{Corresponding author: matthieu.garcin@m4x.org. \newline $^{\text{a}}$ ESILV, 92916 Paris La Défense, France. \newline $^{\text{b}}$ Léonard de Vinci Pôle Universitaire, Research center, 92916 Paris La Défense, France.}} 

\date{\today}

\begin{document}

\maketitle

\begin{abstract}
This paper investigates the impact of COVID-19 on financial markets. It focuses on the evolution of the market efficiency, using two efficiency indicators: the Hurst exponent and the memory parameter of a fractional Lévy-stable motion. The second approach combines, in the same model of dynamic, an alpha-stable distribution and a dependence structure between price returns. We provide a dynamic estimation method for the two efficiency indicators. This method introduces a free parameter, the discount factor, which we select so as to get the best alpha-stable density forecasts for observed price returns. The application to stock indices during the COVID-19 crisis shows a strong loss of efficiency for US indices. On the opposite, Asian and Australian indices seem less affected and the inefficiency of these markets during the COVID-19 crisis is even questionable.
\end{abstract}




\section{Introduction}

The COVID-19 pandemic has strongly affected many persons, either for medical reasons or for the economic aftermath of the various prophylactic measures decided by governments, in particular lockdowns. The evolution of financial markets during the pandemic provides an illustration of the economic impact of these measures. According to several empirical studies, financial markets have indeed been strongly disturbed during this period~\cite{AANH,BBDK,GKL,Pav}. The question of the reaction of financial markets to a crisis is not specific to the COVID-19 pandemic. For example, we can cite a study of the impact of crisis in the 80s and the 90s on a model of dynamic for a stock market~\cite{CRM}. In general, these studies focus on variations of several statistics, such as jump intensity, implied volatility, parameters of factor models, divergence of price return densities, etc. To the best of our knowledge, no paper focuses on measuring the impact of the COVID-19 pandemic on market efficiency.

Market efficiency is the ability of market prices to reflect all the available information, so that no arbitrage is possible. In other words, if markets are efficient, price returns are not correlated with each other and investors are not able to statistically determine what is more profitable between selling and buying a financial asset. Even if this dogma is sometimes questionable in calm periods, we wonder if it can resist to a crisis. We also wonder if we can observe regional disparities, hypothetically related to the magnitude of the outbreak in these regions, and how fast financial markets recover.

The Hurst exponent is a widespread indicator of market efficiency. The long-range feature associated to a Hurst exponent above $1/2$ is indeed traditionally interpreted as indicating predictability of a time series. It is used in finance~\cite{LF,CT,Mitra,BCLS} but also in many other fields such as meteorology, for models of temperature~\cite{DFM}. It is often related to a model of dynamic with a specific fractal property, namely the fractional Brownian motion (fBm), introduced by Mandelbrot and van Ness~\cite{MvN}. The fractal property states that the variance of increments of duration $\tau$ is $\tau^{2H}\sigma^2$, where $H$ is the Hurst exponent and $\sigma^2$ the variance of increments of duration 1. The fBm assumes that increments follow a Gaussian distribution. The fractal property of the fBm is then obtained by introducing a positive (respectively negative) correlation among increments if the Hurst exponent is above (resp. below) $1/2$. In the case where the Hurst exponent is $1/2$, the fBm is simply a standard Brownian motion (Bm), that is with independent increments. In the fBm framework, this value of the Hurst exponent is thus the only one consistent with the efficient market hypothesis (EMH).

However, the approach using the fBm assumes Gaussian price returns. It is not very realistic, since the presence of fat tails in the distribution of price returns is well documented~\cite{LL,BLVW,Nolan2003,GarcinMPRE}. As an alternative to the Gaussian distribution, the alpha-stable distribution is appealing because it includes the Gaussian distribution as a particular case and entails fat tails, whose amplitude is directly related to $\alpha$, a parameter of this distribution. When combining alpha-stable distributions and dependence between the successive price returns, we can get the fractional Lévy-stable motion (fLsm), which is thus a non-Gaussian extension of the fBm~\cite{SamT,ST,SGKMBP}. In this framework, the Hurst exponent is to be decomposed in $m+1/\alpha$, where $m$ is a memory parameter. If $m=0$, adjacent price returns are independent and the market is efficient. We thus use this memory parameter as an alternative efficiency indicator.

The estimation of the fLsm partly relies on the estimation of alpha-stable distributions. Many estimation methods exist for this kind of distribution~\cite{AO,BHW}. In particular, we will focus on McCulloch's method, which is based on empirical quantiles~\cite{McCulloch}. In this paper, we are in addition interested in the dynamic estimation of this distribution, because we want to depict the chronology of the crisis, day after day. Several articles deal with the estimation of time-varying non-parametric densities~\cite{HO,GKL}. The case of parametric densities is simpler as it consists in estimating time-varying parameters. In both the non-parametric and parametric cases, the estimation at a given date takes into account the estimation at the previous date updated by the new observation. The balance between the previous estimation and the new observation is tuned by a discount factor. Several rules are possible for the selection of this free parameter. We focus on the minimization of a criterion related to arguments coming from the field of validation of density forecast~\cite{GKL,DGT}.

In the empirical part of the paper, we study the evolution of the two indicators of market efficiency, $H$ and $m$ for several stock indices, with a significance analysis. We have discovered that the Hurst exponent $H$ detects less often a market inefficiency than does the memory parameter $m$ of an fLsm. We foster the use of this parameter as efficiency indicator instead of the Hurst exponent. It indeed improves the standard Hurst approach insofar as it filters the kurtosis of price returns, which biases the Hurst indicator.

Besides the analysis of the impact of the COVID-19 on market efficiency, the innovative aspects of this paper include a selection rule for the discount factor of a dynamic parametric distribution, an estimation method of dynamic Hurst exponents, and the introduction of the memory parameter of an fLsm as efficiency indicator.

The rest of the paper is organized as follows. Section 2 introduces the estimation of dynamic alpha-stable distributions, along with the selection rule of the discount factor. Section 3 provides some elements on market efficiency and details how indicators are built. Section 4 empirically studies the impact of COVID-19 on market efficiency. Section 5 concludes.

\section{Alpha-stable distributions}

The alpha-stable distribution is a generalization of the Gaussian distribution, appreciated for entailing fat tails. For this reason, it has been widely invoked in signal processing~\cite{SKR,GG2014,GG2016}, with applications for example in medicine~\cite{SGRS,SGKMBP} or in finance~\cite{LL,BLVW,Nolan2003,GarcinMPRE}.

We present below the estimation of a time-varying alpha-stable distribution, which we will apply later in this paper to financial time series. For this purpose, we present first the static estimation as well as the various representations of alpha-stable distributions. Regarding the dynamic distribution, a free parameter is to be selected. It is the discount factor. We propose a rule of selection in the last subsection.

\subsection{Representation}\label{sec:ReprStable}

Four parameters are used to depict a random variable following a stable distribution: $X\sim S_{\alpha}(\gamma,\beta,\mu)$. The parameter $\alpha\in(0,2]$ is the one we will mostly be interested in. It determines the thickness of the tails. The parameter $\beta\in[-1,1]$ is a skewness parameter. If $\alpha=2$ and $\beta=0$, we retrieve the Gaussian distribution. The last two parameters stand for the location ($\mu\in\mathbb R$) and the scale ($\gamma>0$) of the distribution. We do not have any analytic expression for the probability density of $X$, but we can characterize the stable distribution by the mean of its characteristic function:
$$t\mapsto\E\left[\exp\left(itX\right)\right]=\left\{\begin{array}{ll}
\exp\left(i\mu t-\gamma^{\alpha}|t|^{\alpha}\left[1-i\beta \sign(t)\tan\frac{\pi\alpha}{2}\right]\right) & \text{if } \alpha\neq 1 \\
\exp\left(i\mu t-\gamma|t|\left[1+i\beta \sign(t)\frac{2}{\pi}\log|t|\right]\right)  & \text{if }\alpha= 1.
\end{array}\right.$$

We could use the Fourier transform to get the pdf from the characteristic function~\cite{SGRS}, but the above parameterization is not totally satisfactory insofar as the pdf is not continuous in the parameters, in particular when $\alpha=1$~\cite{Nolan,AO}. Indeed, when $\beta>0$, the density is shifted right when $\alpha<1$ and left when $\alpha>1$, with a shift toward $+\infty$ (respectively $-\infty$) when $\alpha$ tends toward 1 by below (resp. above)~\cite{Nolan}. For applications to data and interpretation of the coefficients, this parameterization is thus to be avoided. For this reason, Nolan has proposed to use Zolotarev's (M) parameterization~\cite{Zolotarev}, which is also often called the $S^0$ parameterization. The characteristic function corresponding to $X\sim S^0_{\alpha}(\gamma,\beta,\mu_0)$ is~\cite{Nolan,AO}:
\begin{equation}\label{eq:characS0}t\mapsto\E\left[\exp\left(itX\right)\right]=\left\{\begin{array}{ll}
\exp\left(i\mu_0 t-\gamma^{\alpha}|t|^{\alpha}\left[1+i\beta \sign(t)\tan\frac{\pi\alpha}{2}\left(\gamma^{1-\alpha}|t|^{1-\alpha}-1\right)\right]\right) & \text{if } \alpha\neq 1 \\
\exp\left(i\mu_0 t-\gamma|t|\left[1+i\beta \sign(t)\frac{2}{\pi}\left(\log|t|+\log\gamma\right)\right]\right)  & \text{if }\alpha= 1.
\end{array}\right.
\end{equation}
This alternative parameterization is not far from the $S_{\alpha}$ one. The only difference is about the location parameter, which, in this new setting, corrects the shift exposed above for values of $\alpha$ close to 1:
\begin{equation}\label{eq:mu}
\mu_0=\left\{\begin{array}{ll}
\mu+\beta\gamma\tan\frac{\pi\alpha}{2} &  \text{if } \alpha\neq 1 \\
\mu+\beta\frac{2}{\pi}\gamma\log\gamma  & \text{if }\alpha= 1.
\end{array}\right.
\end{equation}

A Fourier transform makes it possible to get the pdf of a standard variable $S^0_{\alpha}(1,\beta,0)$~\cite{Nolan}:
$$p(x;\alpha,\beta)=\frac{1}{\pi}\int_0^{+\infty}{\cos\left(h(x,t;\alpha,\beta)\right)e^{-t^{\alpha}}dt},$$
with
$$h(x,t;\alpha,\beta)=\left\{\begin{array}{ll}
xt+\beta\tan\frac{\pi\alpha}{2}\left(t-t^{\alpha}\right) & \text{if } \alpha\neq 1 \\
xt+\beta\frac{2}{\pi}t\log t & \text{if }\alpha= 1.
\end{array}\right.$$
We also obtain the pdf of a variable $X\sim S^0_{\alpha}(\gamma,\beta,\mu_0)$ with $\gamma\neq 1$ and $\mu_0\neq 0$, by the mean of a translation, a scaling, and the substitution $s=\gamma t$, starting from the characteristic function provided in equation~\eqref{eq:characS0}:
$$\begin{array}{ccl}
f(x;\alpha,\gamma,\beta,\mu_0) & = & \frac{1}{\pi}\int_0^{+\infty}{\cos\left(h\left(\frac{x-\mu_0}{\gamma},\gamma t;\alpha,\beta\right)\right)e^{-(\gamma t)^{\alpha}}dt} \\
 & = & \frac{1}{\gamma\pi}\int_0^{+\infty}{\cos\left(h\left(\frac{x-\mu_0}{\gamma},s;\alpha,\beta\right)\right)e^{-s^{\alpha}}ds} \\
 & = & \frac{1}{\gamma}p\left(\frac{x-\mu_0}{\gamma};\alpha,\beta\right).
\end{array}$$

The formula of $p$ contains an integral on an unbounded interval. For this reason, other formulations have been proposed~\cite{AO,LC,JMdVG,BHW}. However, in the application to financial series, we find values of $\alpha$ far from 0, so that the truncated integral converges rapidly. We also get the corresponding cdf by numerically integrating $f(x;\alpha,\gamma,\beta,\mu_0)$.

\subsection{Estimation}

Many estimation methods of stable distributions exist~\cite{AO,BHW}. Some of them focus on the sole $\alpha$ parameter, using for instance a regression of extreme quantiles along with the extreme value theory~\cite{HP}. Other methods make it possible to estimate all the parameters. In this class of methods, we can cite the estimation using L-moments, provided that $\alpha>1$~\cite{AO,Hosking}, empirical quantiles~\cite{McCulloch}, the empirical characteristic function~\cite{Press,PHL,Koutrouvelis,KW,SGKMBP}, and the maximum likelihood method~\cite{DuMouchel}. This last method relies on the knowledge of the pdf, which may be approximated as exposed in the previous subsection, using Nolan's work~\cite{Nolan}.

We focus on the method using empirical quantiles, which has been introduced first by Fama and Roll, with the following assumptions: $\alpha>1$, $\beta=0$, and $\mu=0$~\cite{FR}. This method is asymptotically biased. McCulloch proposed an extended version of the method, in which he corrected the asymptotic bias~\cite{McCulloch}. This version is also less restrictive with respect to the parameters, insofar as it only requires to have $\alpha\geq 0.6$.

The McCulloch's method consists in linking the four parameters to five empirical quantiles of probability levels $5\%$, $25\%$, $50\%$, $75\%$, and $95\%$. To this end, we first have to define two intermediate quantities:
$$v_{\alpha}=\frac{Q(0.95)-Q(0.05)}{Q(0.75)-Q(0.25)}$$
and
$$v_{\beta}=\frac{Q(0.95)+Q(0.05)-2Q(0.50)}{Q(0.95)-Q(0.05)},$$
where $Q(p)$ is the theoretical quantile of probability $p$ for a $S_{\alpha}(\gamma,\beta,\mu)$ variable~\cite{McCulloch,AO,BHW}. Neither $v_{\alpha}$ nor $v_{\beta}$ depend on $\gamma$ and $\mu$, so that $\alpha$ and $\beta$ are functions of $v_{\alpha}$ and $v_{\beta}$: $\alpha=\phi_1(v_{\alpha},v_{\beta})$ and $\beta=\phi_2(v_{\alpha},v_{\beta})$. In practice, $\phi_1$ and $\phi_2$ are provided by tables~\cite{McCulloch}. Replacing the theoretical quantiles $Q(p)$ by empirical quantiles $\widehat Q(p)$, we get estimators $\widehat v_{\alpha}$ for $v_{\alpha}$ and $\widehat v_{\beta}$ for $v_{\beta}$, as well as the following estimators for $\alpha$ and $\beta$:
$$\left\{\begin{array}{l}
\widehat \alpha=\phi_1(\widehat v_{\alpha},\widehat v_{\beta}) \\
\widehat \beta=\phi_2(\widehat v_{\alpha},\widehat v_{\beta}).
\end{array}\right.$$
The estimation of $\gamma$ and $\mu$ relies on two other intermediate quantities which only depend on the already estimated $\alpha$ and $\beta$. For simplicity, we introduce the variable $\zeta$ defined by:
\begin{equation}\label{eq:zeta}
\zeta=\left\{\begin{array}{ll}
\mu+\beta\gamma \tan\frac{\pi\alpha}{2} & \text{if } \alpha\neq 1 \\
\mu & \text{if } \alpha=1.
\end{array}\right.
\end{equation}
The intermediate quantities are:
$$v_{\gamma}=\frac{Q(0.75)-Q(0.25)}{\gamma}$$
and
$$v_{\zeta}=\frac{\zeta-Q(0.50)}{\gamma}.$$
They are such that $v_{\gamma}=\phi_3(\alpha,\beta)$ and $v_{\zeta}=\phi_4(\alpha,\beta)$. Their estimators $\widehat v_{\gamma}$ and $\widehat v_{\zeta}$ are obtained by replacing the quantiles by empirical ones. We thus have the following estimators for $\gamma$ and $\zeta$:
$$\left\{\begin{array}{l}
\widehat \gamma=\frac{\widehat Q(0.75)-\widehat Q(0.25)}{\phi_3(\widehat \alpha,\widehat \beta)} \\
\widehat \zeta=\widehat\gamma\phi_4(\widehat \alpha,\widehat \beta)+\widehat Q(0.50).
\end{array}\right.$$
The deduction of the estimator of $\mu$ from $\widehat \zeta$ is straightforward using equation~\eqref{eq:zeta}, as well as the version $\mu_0$ of the location parameter in the parameterization $S^0$ using equation~\eqref{eq:mu}. In particular $\widehat \mu_0 = \widehat \zeta$ as soon as $\alpha\neq 1$.


\subsection{Dynamic estimation}\label{sec:DynEst}

The time-varying adaptation of McCulloch's estimation technique amounts to estimating time-varying quantiles. Indeed, if we are able to infer dynamic quantiles $Q_t(p)$, the first two McCulloch statistics are defined in a dynamic fashion:
$$\left\{\begin{array}{ccl}
v_{\alpha,t} & = & \frac{Q_t(0.95)-Q_t(0.05)}{Q_t(0.75)-Q_t(0.25)} \\
v_{\beta,t} & = & \frac{Q_t(0.95)+Q_t(0.05)-2Q_t(0.50)}{Q_t(0.95)-Q_t(0.05)} ,
\end{array}\right.$$
so that we get time-varying $\alpha$ and $\beta$ parameters. Time-varying quantiles make it also possible to define the last two McCulloch statistics and finally to fully estimate a dynamic alpha-stable probability distribution.

The subject of estimating dynamic quantiles is largely handled by the econometric literature. The favoured appraoch is based on quantile autoregression~\cite{KX,DH,GCC}, like in the application to value-at-risk known as CAViaR~\cite{EM}. A drawback of quantile regression is that different quantiles may cross: the monotonicity of quantiles is not necessarily preserved. The dynamic additive quantile, while keeping the autoregressive approach, deals with this limitation~\cite{GJ}. 

All the quantile autoregressions are based on a model of dynamic and we prefer to introduce a method in which we do not specify the evolution of quantiles with respect to previously estimated ones. The inspiration of such a model-free approach comes from the non-parametric statistics literature, in which we can estimate for example time-varying moments or even time-varying probability densities~\cite{HO}. In this perspective, we simply consider that the price returns $X_t$ are alpha-stable random variables, each following its own alpha-stable distribution $F_t$, defined by parameters $\alpha_t$, $\beta_t$, $\omega_t$, and $\mu_t$. In each distribution $F_t$, we only observe one variable $X_t$. This is not enough for estimating $F_t$. For this reason, we add another assumption on the dynamic of this time-varying distribution. We indeed consider that it evolves smoothly, so that price returns close in time will follow a close distribution. 

Thanks to this assumption, we can use several price returns at times close to $t$ in order to estimate $F_t$ or its related quantiles $Q_t(p)$. We could estimate these quantiles $Q_t(p)$ by their empirical versions. For example, if one considers 100 observations, the empirical quantile of probability $p$ is the $\lceil 100 p\rceil$-th smallest of the 100 observations. In other words, each of the 100 observations is associated with a probability $1/100$. However, price returns closer in time induce a lower bias in the estimation of $Q_t(p)$, so that the probability associated with these recent returns in the cdf related to $Q_t$ should be higher. Thus, we use an exponentially-weighted quantile estimator (EWQ) and the corresponding discrete probability $\widehat P_t^{EWQ}$, such that $\widehat P_t^{EWQ}(x)$ is the probability for the price return at time $t$ to be equal to $x$~\cite{NM}. To this end, we estimate the EWQ-style discrete probability function by assigning to each observation $X_i$, for $i\in\llbracket 1,t\rrbracket$, a probability $p^{\omega}_{t,i}$ depending on a discount factor $\omega\in(0,1)$, so that $\widehat P_t^{EWQ}(X_i)=p^{\omega}_{t,i}$. The probability $p^{\omega}_{t,i}$ follows a standard form of exponential weight:
$$p^{\omega}_{t,i}=\frac{1-\omega}{1-\omega^t}\omega^{t-i}.$$
This expression ensures that the probabilities sum to 1, as required for probabilities:
$$\sum_{i=1}^t{p^{\omega}_{t,i}}=1.$$ 
For $\omega\neq 1$ and big values of $t$, $\omega^t$ is close to zero and we have the approximation $p^{\omega}_{t,i}\approx (1-\omega)\omega^{t-i}$. In addition to the discrete probability $\widehat P_t^{EWQ}$, we can define the cdf $\widehat F_t^{EWQ}$ associated with the EWQ:
\begin{equation}\label{eq:cdfEWQ}
\widehat F_t^{EWQ}(x)=\sum_{i=1}^{t}{\widehat P_t^{EWQ}(X_i)\indic_{X_i\leq x}}=\sum_{i=1}^{t}{p^{\omega}_{t,i}\indic_{X_i\leq x}}.
\end{equation}
We can also write recursively this cdf~\cite{NM}: 
$$\widehat F_t^{EWQ}(x)=\omega\widehat F_{t-1}^{EWQ}(x)+(1-\omega)\indic_{X_t\leq x}.$$
More precisely, for each new observation $X_t$, the EWQ-style cdf associated with each past observations is discounted at the constant rate $\omega$, whereas the new observation is provided with the highest probability, $1-\omega$. The rationale is the following: the more recent the observation, the more likely its future occurrence.

These discrete probabilities make it possible to calculate easily time-varying quantiles, as generalized inverse functions of $\widehat F_t^{EWQ}$. Compared to quantiles determined from dynamic kernel densities, which suffer from a high algorithmic complexity~\cite{YJ,HO}, the EWQ approach seems computationally more performing. Indeed, for extreme quantiles, the algorithmic complexity of our method is less than linear with respect to the number of observations considered at each date. 

We now describe precisely the procedure for calculating the EWQs. We begin with a first estimation, at date $t_0$, of the quantile of probability $p$, $\widehat Q^{\omega}_{t_0}(p)$, using the EWQ approach. To estimate the quantiles, we build the matrix $M^{\omega}_{t_0}$ containing the past observed price returns till $t_0$, $\{X_i\}_{i\leq t_0}$, sorted in descending order, along with their corresponding historical EWQ-style probability:
$$M^{\omega}_{t_0}=\left(\begin{array}{cc}
X_{\pi_{t_0}(t_0)} & p^{\omega}_{t_0,\pi_{t_0}(t_0)} \\
\vdots & \vdots \\
X_{\pi_{t_0}(i)} & p^{\omega}_{t_0,\pi_{t_0}(i)} \\
\vdots & \vdots \\
X_{\pi_{t_0}(1)} & p^{\omega}_{t_0,\pi_{t_0}(1)}
\end{array}\right),$$
where $X_{\pi_{t_0}(i)}$ is the $i$-th order statistic among the $t_0$ first observations and is obtained with the help of a permutation $\pi_{t_0}$: $\underset{i\leq t_0}{\min}\{X_i\}=X_{\pi_{t_0}(1)}\leq X_{\pi_{t_0}(2)} \leq ...\leq X_{\pi_{t_0}(t_0)}=\underset{i\leq t_0}{\max}\{X_i\}$. Thanks to the definition of the generalized quantile and to equation~\eqref{eq:cdfEWQ}, we get the following estimator for the quantile of probability $p$:
\begin{equation}\label{eq:QuantileIter}
\widehat Q^{\omega}_{t_0}(p)=X_{\pi_{t_0}\left(\min\left\{\tau\in\{1,...,t_0\}\left|\sum_{i=1}^{\tau}{p^{\omega}_{t_0,\pi_{t_0}(i)}}\geq p\right.\right\}\right)},
\end{equation}
which is simply the lowest price return $X$ such that the cumulated probability associated to lower returns, $\widehat F_{t_0}^{EWQ}(X)$, reaches $p$.

Iteratively, we can update this quantile in the following manner. We suppose we are given the probability distribution till time $t-1$ and that we want to estimate a quantile at time $t$. We first apply a probability decay to $M^{\omega}_{t-1}$ by a simple matrix product, so that we get a new matrix $\widetilde M^{\omega}_{t-1}$ containing sorted past observations till time $t-1$ along with their new probability at time $t$:
$$\widetilde M^{\omega}_{t-1}=M^{\omega}_{t-1}\left(\begin{array}{cc}
1 & 0 \\
0 & \omega
\end{array}\right).$$
Then, we insert the new observation of time $t$ in the matrix $\widetilde M^{\omega}_{t-1}$ thanks to a binary search in its first column. In the inserted line, we write the corresponding probability $1-\omega$ in the second column. If we write $\mathcal I_{t}$ the position of the new observation, we get the new probability distribution matrix $M^{\omega}_{t}$ by:
$$M^{\omega}_{t}=\left(\begin{array}{cc}
X_{\pi_{t}(t)} & p^{\omega}_{t,\pi_{t}(t)} \\
\vdots & \vdots \\
X_{\pi_{t}\left(\mathcal I_{t}+1\right)} & p^{\omega}_{t,\pi_{t}\left(\mathcal I_{t}+1\right)} \\
X_{\pi_{t}\left(\mathcal I_{t}\right)} & p^{\omega}_{t,\pi_{t}\left(\mathcal I_{t}\right)} \\
X_{\pi_{t}\left(\mathcal I_{t}-1\right)} & p^{\omega}_{t,\pi_{t}\left(\mathcal I_{t}-1\right)} \\
\vdots & \vdots \\
X_{\pi_{t}(1)} & p^{\omega}_{t,\pi_{t}(1)}
	\end{array}\right)=\left(\begin{array}{cc}
X_{\pi_{t-1}(t-1)} & \omega p^{\omega}_{t-1,\pi_{t-1}(t-1)} \\
\vdots & \vdots \\
X_{\pi_{t-1}\left(\mathcal I_{t}\right)} & \omega p^{\omega}_{t-1,\pi_{t-1}\left(\mathcal I_{t}\right)} \\
X_{t} & 1-\omega \\
X_{\pi_{t-1}\left(\mathcal I_{t}-1\right)} & \omega p^{\omega}_{t-1,\pi_{t-1}\left(\mathcal I_{t}-1\right)} \\
\vdots & \vdots \\
X_{\pi_{t-1}(1)} & \omega p^{\omega}_{t-1,\pi_{t-1}(1)}
\end{array}\right).$$
We can then calculate the empirical quantile in a manner similar to equation~\eqref{eq:QuantileIter}:
$$\widehat Q^{\omega}_{t}(p)=X_{\pi_{t}\left(\min\left\{\tau\in\{1,...,t\}\left|\sum_{i=1}^{\tau}{p^{\omega}_{t,\pi_{t}(i)}}\geq p\right.\right\}\right)}.$$

In order to diminish the algorithmic complexity of this method, if we are looking for a quantile above the probability $0.5$, we prefer the following definition of the quantile, which is mathematically consistent with the one provided above:
$$\widehat Q^{\omega}_{t}(p)=X_{\pi_{t}\left(\min\left\{\tau\in\{1,...,t\}\left|\sum_{i=\tau+1}^{t}{p^{\omega}_{t,\pi_{t}(i)}}\leq 1-p\right.\right\}\right)}.$$

Finally, using McCulloch's method, these time-varying estimations of quantiles make it possible to estimate the dynamic parameters of the stable distribution, that we note $\widehat\alpha^{\omega}_t$, $\widehat\beta^{\omega}_t$, $\widehat\gamma^{\omega}_t$, and $\widehat\mu^{\omega}_t$. We stress the fact that the EWQ method is only intended to estimate non-parametric quantiles in order to infer time-varying parameters. In particular, in what follows, the time-varying cdf on which we focus is not the non-parametric EWQ-style cdf but the parametric alpha-stable cdf with the time-varying parameters estimated above.

\subsection{Selection of the discount factor}\label{sec:discount}

The above dynamic estimation of quantiles and of the parameters of stable distributions relies on a free parameter, the discount factor $\omega$, which depicts how fast the dynamic distribution evolves. If $\omega$ is close to 1, the distribution is almost constant. If $\omega$ is lower, the evolution of the distribution is faster and the description of the last observations will be more accurate. Nevertheless, this accuracy may be excessive and the evolution of the distribution may be non-significant. Indeed, in the extreme situation where $\omega$ is close to zero, the distribution will be very narrow and centered on the last observation, with a big divergence between two successive distributions. In order to find a good balance between accuracy and robustness, we decide to select the $\omega$ maximizing the ability of the density $\widehat f^{\omega}_t$ estimated at date $t$ to forecast the density of $X_{t+1}$, the price return at time $t+1$.

Several definitions of what a good density forecast is are possible. Indeed, the true density at time $t+1$ is never observed, so that we can only rely on one observation drawn in this density. In the non-parametric literature about time-varying densities, we find for instance a selection rule for the free parameter of the densities based on the maximization of a likelihood criterion~\cite{HO}. We think that this criterion does not take properly into account the possibility of the occurrence of extreme events. The alternative solution we follow is based on an adaptation of a method coming from the literature of density forecast evaluation~\cite{GKL}. Indeed, even if $\widehat f^{\omega}_t$ varies with $t$, so that we are provided with only one observation in this distribution, a simple transformation of each price return defines a distribution which remains the same through time. This transformation is the probability integral transform (PIT), usually introduced in the perspective of density forecast evaluation~\cite{DGT}:
\begin{equation}\label{eq:PIT}
Z^{\omega}_t=\widehat F^{\omega}_{t-1}(X_t),
\end{equation}
where $\widehat F^{\omega}_{t-1}$ is the cdf corresponding to $\widehat f^{\omega}_{t-1}$. In this literature, two conditions are required for the PITs: the $Z^{\omega}_t$ must follow a uniform distribution in $[0,1]$ and they must be independent from each other. 

The translation of these rules to the selection of free parameters in density estimation leads to two properties regarding $\omega$~\cite{GKL}:
\begin{itemize}
\item uniformity of the PITs $Z^{\omega}_{t_0},...,Z^{\omega}_T$: $\omega$ is to be selected so as to minimize the divergence between their empirical distribution and a uniform distribution,
\item independence of the PITs: $\omega$ is to be selected so as to minimize the discrepancy, that is, for each subinterval of $[t_0,T]$ of size greater than a threshold $\nu$, the divergence between the empirical distribution of the corresponding PITs and a uniform distribution is to be minimized.
\end{itemize}
In the perspective of the selection of a free parameter of a time-varying distribution, we can define the above divergence as a Kolmogorov-Smirnov statistic with an adaptation to compare directly divergences of distributions estimated on samples of different sizes~\cite{GKL}. Moreover, the minimal size $\nu$ of the subintervals considered is intended to be a threshold above which the asymptotic framework required by the Kolmogorov-Smirnov statistic is satisfied. We consider $\nu=22$ days, so that we expect the PITs to be uniform for scales larger than one month. As a consequence, the criterion to be minimized is:
$$d_{\nu}(Z^{\omega}_{t_0+1},...,Z^{\omega}_T)=\underset{t_0+1\leq s<s+\nu-1\leq t\leq T}{\max}\left(\sqrt{t-s+1}\times k(Z^{\omega}_s,...,Z^{\omega}_t)\right),$$
in which $k$ is the standard Kolmogorov-Smirnov statistic with respect to the uniform distribution:
\begin{equation}\label{eq:KS}
k(Z^{\omega}_{s},...,Z^{\omega}_t)=\underset{s\leq u\leq t}{\max}\left|\frac{u-s}{t-s}-Z^{\omega}_{\rho(u;s,t)}\right|,
\end{equation}
where $v\in\llbracket a,b\rrbracket\mapsto\rho(v;a,b)$ is a permutation of $\llbracket a,b\rrbracket$ defining the new order $Z^{\omega}_{\rho(a;a,b)}\leq Z^{\omega}_{\rho(a+1;a,b)}\leq...\leq Z^{\omega}_{\rho(b-1;a,b)}\leq Z^{\omega}_{\rho(b;a,b)}$~\cite{GKL}. In equation~\eqref{eq:KS}, $(u-s)/(t-s)$ is the empirical cdf of the PITs, whereas the sorted PIT $Z^{\omega}_{\rho(u;s,t)}$ is the theoretical uniform cdf. Finally, the optimal discount factor is defined as the solution of the following equation:
\begin{equation}\label{eq:omega}
\omega^{\star}=\underset{0<\omega< 1}{\operatorname{argmin}} \text{ } d_{\nu}(Z^{\omega}_{t_0+1},...,Z^{\omega}_T).
\end{equation}

We can easily apply the above method to the case of an alpha-stable distribution. We only have to pay attention to the definition of the PIT in equation~\eqref{eq:PIT}. Indeed, it relies on the cdf, which does not follow a simple expression for alpha-stable distributions. Nevertheless, as exposed in Subsection~\ref{sec:ReprStable}, numerical methods make it possible to calculate both the pdf and the cdf of an alpha-stable distribution. Therefore, the estimated cdf $x\mapsto\widehat F^{\omega}_{t-1}(x)$ in equation~\eqref{eq:PIT} is the numerically evaluated cdf of an alpha-stable distribution with estimated parameters: $x\mapsto F(x;\widehat\alpha^{\omega}_{t-1},\widehat\gamma^{\omega}_{t-1},\widehat\beta^{\omega}_{t-1},(\widehat\mu_0)^{\omega}_{t-1})$.

\section{Market efficiency}

The market efficiency is a usual assumption in finance, which is in particular invoked when pricing derivatives. The EMH states that asset price series follow a random walk~\cite{Fama}. Investors and market makers update their expectations and their quotes at each instant, using the available information, so that it is not possible to beat the market. The EMH is convenient, because working with independent price returns makes the financial mathematics easier. But practitioners know that the EMH is not very realistic. The asset management industry in general aims at performing statistical arbitrages, whatever the transaction time scale, whether it is less than a second or more than a month. Maybe the asset manager will not win at each time, but in average he should win. Standard models relying on the EMH, such as the widespread geometric Brownian motion, are not consistent with the existence of statistical arbitrages.

We can also stress other unrealistic components of the EMH, such as the availability of the same information for every investor and market maker, as well as the rationality of all the agents, or the fact that we can substitute two assets provided that they are equally risky. Several alternatives to the EMH aim at describing more realistically the financial markets. It is the case of the fractal market hypothesis~\cite{Peters} and of the adaptive market hypothesis (AMH)~\cite{Lo}. The AMH is a good compromise between the standard EMH and the reality of statistical arbitrage. It states that a model can predict the market in average. But such a model can provide investors with performing forecasts during a limited time only, because other investors and market makers will progressively adapt their own models and decisions to this performing model. The AMH thus leads to a long-term efficiency of the market and allows statistical arbitrages for small time scales only.

Besides, it is worth noting that the price series of some assets are not far from the EMH. It is the case for major stock indices. But empirical studies show that these indices have fluctuating efficiency and may encounter a loss of efficiency during financial crisis~\cite{AVE,LBK}. Since the decisions of market makers and investors may not be the same if markets are efficient or if they aren't, it is important for them to determine from time series of prices whether the markets are efficient or not. We can cite many statistical indicators of market efficiency~\cite{RSVD}. Some of them look for a predictability of price returns~\cite{CRS}, using for example the amplitude of the parameters of a time-varying AR model~\cite{INW,Noda}. Other indicators measure a deviation from a random walk, using for example variance ratios~\cite{CD} or a combination of several statistics such as fractal dimension and entropy~\cite{KV,Kristoufek,FYY}. But the most widespread indicator of market efficiency seems to be the Hurst exponent~\cite{LF,CT,Mitra,Garcin2017}.

The Hurst exponent is an indicator of long-range memory. Mandelbrot and van Ness introduced the fBm as a model consistent with a given Hurst exponent and extending the standard Bm by making the increments dependent on each other~\cite{MvN}. For this reason, the fBm is a popular model in finance, provided that practitioners do not want to comply with the EMH. In the fBm, the Hurst exponent $H$ is also linked to the fractal property of the series, insofar as the variance of the increments of duration $\tau$ is $\tau^{2H}$ times the variance of increments of duration 1. Some estimators of the Hurst exponent use this fractal property instead of the long-range memory. However, given the fractal property estimated on the dataset, the fBm is not the only possible model. Other models may indeed have the same estimated Hurst exponent as the fBm but, relying on another specification, they may lead to other conclusions regarding the dependence of the increments and thus regarding the efficiency of the markets~\cite{AP}. This fact has been documented for instance for foreign exchange rates, for which the stationarity of the time series biases the estimation of the Hurst exponent in the perspective of an fBm~\cite{GarcinLamperti,GarcinEstimLamp}. Indicators of market efficiency should not bypass the diversity of models featuring a fractal property and consistent with the estimated Hurst exponent. Extensions of the fBm include some specificities for the Hurst exponent: it may vary deterministically through time, as in the multifractional Brownian motion~\cite{PLV,BJR,Coeurjolly,Garcin2017}, it may be a random process, as in the multifractional process with random exponent~\cite{AT,BP2011,Frezza,GarcinMPRE}, or it may even be asymmetric~\cite{CCX,SapinaADFA}. 

In what follows, after a presentation of the Hurst exponent in the perspective of the fBm, we will focus on a model in which increments are not necessarily Gaussian. In this framework, the fLsm is a natural extension of the fBm, in which increments follow an alpha-stable distribution~\cite{ST,WBMW,SGKMBP,GarcinMPRE}. The fractal property of this process takes into account both the dependence between increments and the tail parameter of the alpha-stable distribution. This model thus makes it possible to disentangle the kurtosis of price returns and the efficiency of the market. It provides us with richer information than the sole Hurst exponent. This refinement is not superfluous. We will indeed see in the empirical section that the conclusions regarding the efficiency of stock indices are not the same if we consider the Hurst exponent or the fLsm-based indicator of market efficiency.

\subsection{Hurst exponents}\label{sec:Hurst}

The Hurst exponent was originally introduced by Harold Edwin Hurst as an indicator of long-range memory that could be obtained thanks to the rescaled range (R/S) analysis~\cite{Hurst}. Later, alternative estimation methods appeared, such as the detrended fluctuation analysis~\cite{PBHSSG}, or the absolute-moment method~\cite{BJR,BCI,Bianchi,Garcin2017}, which is related to the notion of generalized Hurst exponent (GHE)~\cite{BV,DM}. The absolute-moment method is a method mainly used by the community of statisticians of stochastic processes, because, contrary to the R/S analysis, it is strongly related to a stochastic process, namely the fBm.

The fBm is a generalization of the standard Bm. Increments of the fBm are Gaussian, since the fBm is a fractional integral or a fractional derivative of a Bm, $W_t$~\cite{MvN}:
$$B_H(t)=\frac{\sigma}{\Gamma\left(H+\frac{1}{2}\right)}\int_{-\infty}^{\infty}{\left((t-s)_+^{H-1/2}-(-s)_+^{H-1/2}\right)dW(s)}.$$
The two parameters of the fBm are the Hurst exponent $H$, and the volatility parameter $\sigma$. If $H=1/2$, the fBm  is a Bm. If $H>1/2$ (respectively $H<1/2$), the fBm is the fractional integral (resp. fractional derivative) of order $H-1/2$ (resp. $1/2-H$) of a Bm; increments are thus positively (resp. negatively) correlated. 

The absolute-moment method uses another definition of the fBm, which is consistent with the integral form provided above. Indeed, an fBm is also the only zero-mean Gaussian process, with zero at the origin, such that, for $s,t\geq 0$:
$$\E\{B_H(t)B_H(s)\}=\frac{\sigma^2}{2}(|t|^{2H}+|s|^{2H}-|t-s|^{2H}).$$
From this covariance, we get the self-similarity property: 
$$\E\{|B_H(t)-B_H(s)|^k\}=\sigma^2|t-s|^{kH},$$
for $k>0$. This property states that the $k$-order absolute moment of the increments of duration $|t-s|$ is proportional to $|t-s|^{kH}$. Comparing two scales thus makes it possible to estimate $H$. If we focus on the two smallest scales, we indeed get the following estimator, for a time series of log-prices $X_1, X_2,...,X_t$:
$$\widehat H_k =\frac{1}{k}\log_2\left(\frac{(t-1)\sum_{i=1}^{t-2}{|X_{i+2}-X_i|^k}}{(t-2)\sum_{i=1}^{t-1}{|X_{i+1}-X_i|^k}}\right),$$
which converges almost surely toward $H$~\cite{BJR,BCI}.

We could use this estimator of $H$ as an indicator of market efficiency. But, we are mostly interested here in evaluating the evolution of market efficiency through time. We need therefore a time-varying version of this estimator. This question is not new and the solution put forward in the literature is often based on the estimation in sliding windows~\cite{Coeurjolly,BP2010}, possibly with a smoothing of the raw series of Hurst exponents as a post-processing~\cite{Garcin2017}. But we prefer a method more consistent with the smoothing applied above for estimating the parameters of a distribution, that is with an exponential weighting, insofar as it overweights more recent observations and is thus a more relevant picture of the current state of the market. The closest method in the literature is a time-varying GHE using the exponentially-weighted moving average~\cite{MMGA}. The main difference with the method we propose is that the GHE is always based on a linear regression of log-absolute moments of increments on several log time scales. Contrary to the GHE, we focus on only two scales, so that we get a simpler closed-form estimator:
$$\widehat H^{\omega}_{k,t} =\frac{1}{k}\log_2\left(\frac{(t-1)\sum_{i=1}^{t-2}{\omega^{t-2-i}|X_{i+2}-X_i|^k}}{(t-2)\sum_{i=1}^{t-1}{\omega^{t-1-i}|X_{i+1}-X_i|^k}}\right).$$
Beyond this simple formula, a more efficient implementation method is possible. Given the one-step statistic $M_{k,1,t}=\sum_{i=1}^{t-1}{\omega^{t-1-i}|X_{i+1}-X_i|^k}$ and the two-step statistic $M_{k,2,t}=\sum_{i=1}^{t-2}{\omega^{t-2-i}|X_{i+2}-X_i|^k}$, it consists of the following recurrence:
\begin{equation}\label{eq:DynEstH}
\left\{\begin{array}{ccl}
M_{k,1,t+1} & = & \omega M_{k,1,t}+|X_{t+1}-X_t|^{k} \\
M_{k,2,t+1} & = & \omega M_{k,2,t}+|X_{t+1}-X_{t-1}|^{k} \\
\widehat H^{\omega}_{k,t+1} & = & \frac{1}{k}\log_2\left(\frac{tM_{k,2,t+1}}{(t-1)M_{k,1,t+1}}\right).
\end{array}\right.
\end{equation}

We will discuss in Section~\ref{sec:fLsm} the selection of the parameter $k$ in equation~\eqref{eq:DynEstH}. The question of the optimal choice of the discount factor $\omega$ in the estimator $\widehat H^{\omega}_{k,t}$ is also to be addressed. We could imagine to adapt the method exposed in section~\ref{sec:discount} to the case of an fBm. But it sounds more relevant to use the same discount factor for all the statistics of our work. We will thus simply use the discount factor chosen for estimating the time-varying stable distribution.

\subsection{Fractional Lévy-stable motion}\label{sec:fLsm}

The fLsm is a generalization of the fBm in which increments follow an alpha-stable distribution, which admits the Gaussian distribution as a particular case. The fLsm is defined as the fractional integral or fractional derivative of a symmetric Lévy-stable motion~\cite{SamT,ST,SGKMBP}:
$$B_{H,\alpha}(t)=\frac{1}{C_{H,\alpha}}\int_{-\infty}^{\infty}{\left((t-s)_+^{H-1/\alpha}-(-s)_+^{H-1/\alpha}\right)dL_{\alpha,\gamma}(s)},$$
where $L_{\alpha,\gamma}(t)$ is a symmetric $\alpha$-stable process of scale parameter $\gamma$ and
$$C_{H,\alpha}=\left(\int_{\mathbb R}{\left|(t-s)_+^{H-1/\alpha}-(-s)_+^{H-1/\alpha}\right|^{\alpha}ds}\right)^{1/\alpha}.$$ 
In other words, increments follow a symmetric stable law and, if and only if $H-1/\alpha>0$, non-overlapping increments are positively dependent, that is with a positive codifference or a positive covariation~\cite{LT}. The parameter $H$ controls the scaling behaviour of the process, in the same manner as in the fBm, and the parameter $\alpha$ controls the thickness of the tails. The lower $\alpha$, the fatter the tails. If $\alpha=2$, increments are Gaussian and the fLsm is an fBm. When comparing an fBm and an fLsm, the fractal feature of the latter is not obtained only by adjusting the dependence of the increments but, in addition, by tuning the kurtosis of the underlying law. Therefore, we can write the Hurst exponent as the combination of a tail component, $1/\alpha$, and a memory parameter, $m$:
\begin{equation}\label{eq:Hmalpha}
H=\frac{1}{\alpha}+m.
\end{equation}
In this framework, the Hurst exponent is not the most relevant indicator of market efficiency and one should instead use $m$. As we have proposed time-varying estimators both for $\alpha$ and for $H$, we write the following time-varying estimator for $m$:
\begin{equation}\label{eq:Hmalphat}
\widehat{m}^{\omega}_t=\widehat H^{\omega}_{k,t}-\frac{1}{\widehat\alpha^{\omega}_t},
\end{equation}
where $\widehat\alpha^{\omega}_t$ is the estimator of the parameter of the stable law defined in section~\ref{sec:DynEst}. Efficient markets then correspond to $\widehat{m}^{\omega}_t$ close to zero. The multifractional multistable motion is a model allowing this kind of dynamic with smoothly time-varying parameters $\alpha_t$ and $\gamma_t$. It is a localisable process in the sense that such a process admits at each time a local form, also called tangent process, which in our case is an fLsm~\cite{FLL,LG}.

We also have to select an appropriate $k$ for the estimator of $H$ in equation~\eqref{eq:Hmalphat}. A widespread choice is $k=2$, because it minimizes the variance of the estimator. However, for $\alpha$-stable variables, the absolute moments are only defined for $k<\alpha$~\cite{KSS}. We can mitigate this theoretical constraint by noting that the statistics of equation~\eqref{eq:DynEstH} are finite even for higher values of $k$. In the empirical part of this paper, we first follow the widespread choice $k=2$. Then, we confirm the results with a lower value of $k$, namely $k=0.5$, which is consistent with the minimal estimated $\alpha$ appearing in Figure~\ref{fig:alpha}.

\section{Empirical study}

 
We apply the above method to ten stock indices of various regions: USA (S\&P 500, S\&P 100), Europe (EURO STOXX 50, Euronext 100, DAX, CAC 40), Asia (Nikkei, KOSPI, SSE 180), and Australia (S\&P/ASX 200). We have used data from Yahoo finance in the time interval between the 1st May 2015 and the 29th June 2020. The first date at which we estimate stable densities and the parameters of an fBm and an fLsm, that is $t_0$, is the 1st November 2019. The period of study includes the financial crisis sparked by the COVID-19 pandemic.

We first determine for each stock index the optimal discount factor $\omega$ as in equation~\eqref{eq:omega}. The results are displayed in Table~\ref{tab:optimal}. We observe that the optimal discount factors are close to 0.95, whatever the index considered. For the rest of the empirical study, we consider a common discount factor $\omega^m$, so that we can make fair comparisons between stock indices. We choose $\omega^m=0.956$, which is the highest optimal discount factor measured for the various stock indices. This conservative choice limits the risk of providing a new observation with a too high weight in the dynamic estimators. A lower value could lead to spurious conclusions regarding the market efficiency for some stock indices.

\begin{table}[htbp]
\centering
\begin{tabular}{|l|c|}
\hline
Stock index & $\omega^{\star}$ \\
\hline
S\&P 500 & 0.946  \\
S\&P 100 &  0.939  \\
EURO STOXX 50 & 0.949 \\
Euronext 100 & 0.954 \\
DAX &  0.949 \\
CAC 40 & 0.952 \\
Nikkei 225 & 0.951 \\
KOSPI & 0.952 \\
SSE 180 & 0.946 \\
S\&P/ASX 200 & 0.956 \\
\hline
Mean value & 0.949 \\
Median value & 0.950 \\
\hline
\end{tabular}
\begin{minipage}{0.7\textwidth}\caption{Optimal discount factor $\omega^{\star}$ for several stock indices for stable densities between November 2019 and June 2020.}
\label{tab:optimal}
\end{minipage}
\end{table}

\begin{figure}[htb]
	\centering
		\includegraphics[width=0.45\textwidth]{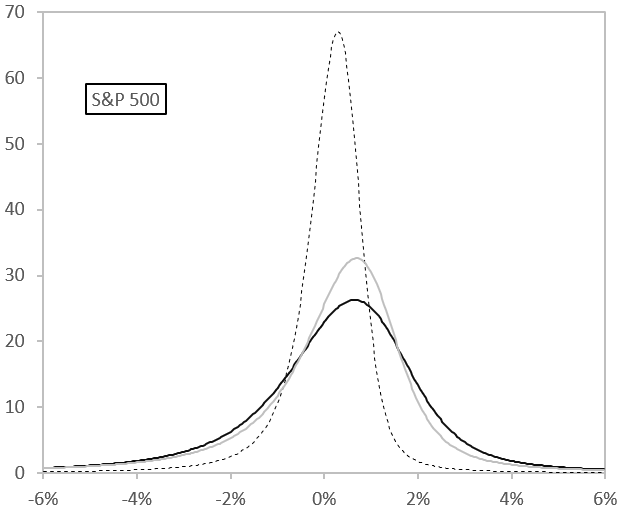} 
		\includegraphics[width=0.45\textwidth]{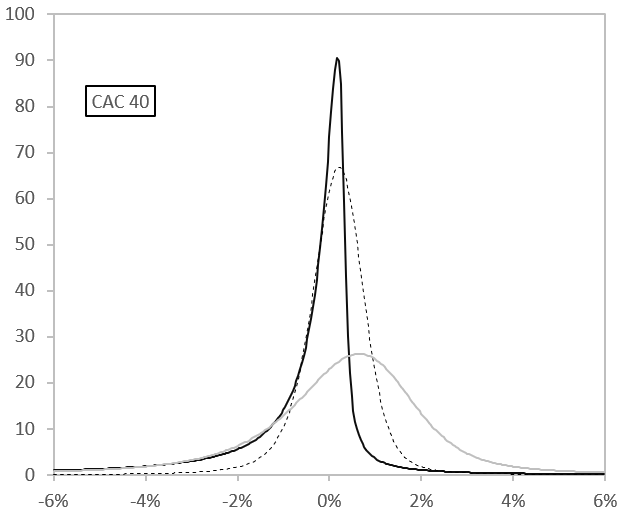} \\
		\includegraphics[width=0.45\textwidth]{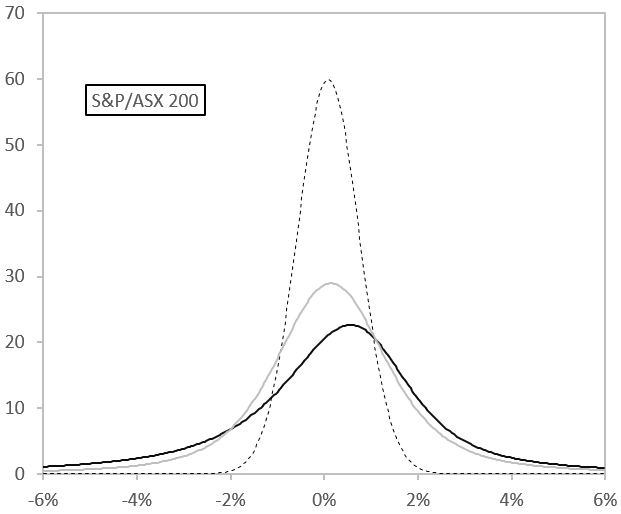} 
		\includegraphics[width=0.45\textwidth]{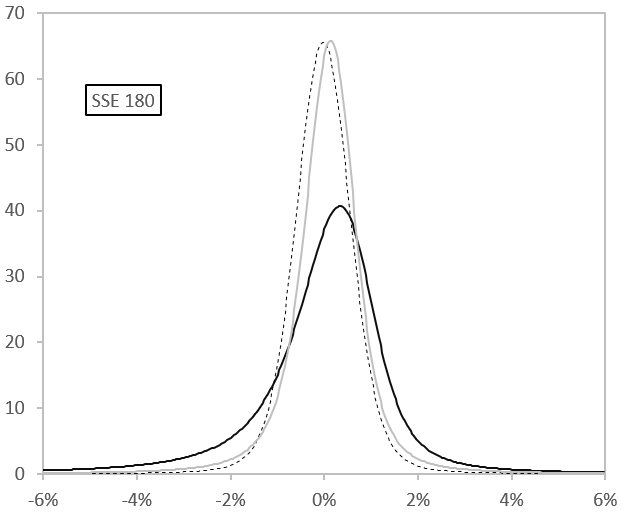} 
\begin{minipage}{0.7\textwidth}\caption{Estimated dynamic stable density of daily price returns for S\&P 500 (top left), CAC 40 (top right), S\&P/ASX 200 (bottom left), and SSE 180 (bottom right) indices. The dotted line is the density at the 5th November 2019, the black continuous line at the peak of the crisis, the grey line at the 29th June 2020.}
	\label{fig:pdf}
\end{minipage}
\end{figure}

Using $\omega^m$, we are able to determine the dynamic pdf of daily price returns. We display in Figure~\ref{fig:pdf} these densities for four stock indices corresponding to regions with a different timing in the growth of the outbreak: S\&P 500, the French CAC 40, the Chinese SSE 180, and the S\&P/ASX 200 indices. For each of these indices, we plot the pdf before the crisis, in November 2019, at the peak of the crisis, and a the end of our sample, late June 2020. The peak of the crisis is not the same for all the indices. We define the peak date as the one leading to the maximal value for $|m|$. This date is in March for the four indices. Exact dates are provided in Table~\ref{tab:rangeM}. For the four indices, the pdf at the peak shows fat tails and asymmetry. Late June, the pdf still has these features, except for SSE 180 index, for which the pdf is very similar to the one before the crisis, indicating a very fast recovery in China. The case of CAC 40 at the peak is also of interest, because the advent of fat tails and asymmetry is so abrupt that it does not crush the body of the pdf, contrary to other indices. 

In this paper, we are mostly interested in determining whether the financial markets are efficient during a financial crisis. For this purpose, we have introduced two indicators. The first one is the widespread Hurst exponent, estimated here in a dynamic fashion as exposed in Section~\ref{sec:Hurst}. But the Hurst exponent $H$ provided above is an indicator of dependence between price returns only if  these price returns are Gaussian. In a more general framework, if we consider the possibility of fat tails by the mean of an alpha-stable distribution, we define another efficiency indicator by the memory parameter $m$ of an fLsm, as exposed in Section~\ref{sec:fLsm}. In the first approach, the market is efficient for $H=1/2$. In the second approach, which is more accurate because it takes into account the kurtosis of price returns thanks to the $\alpha$ parameter, the market is efficient for $m=0$. The null hypothesis $H_0$ is the efficiency of each market. In order to know for which threshold of efficiency indicator we can reject $H_0$ with a given confidence $p$, we perform a simulation. We consider that the right price model corresponding to $H_0$ is a geometric Bm. We thus simulate a time series following this model. The length of the simulated series used for estimating the first parameters in $t_0$ is the same as in our financial dataset. We then simulate 4,000 other dates. For each of these dates, we estimate $H$, $m$, and $\alpha$ dynamically, using the discount factor $\omega^m$. We consider that the bounds of the confidence interval with confidence level $p$, for the estimated $H$, $m$, and $\alpha$, are the empirical quantiles of the corresponding parameters, estimated on the simulations, for probabilities $(1-p)/2$ and $(1+p)/2$.

\begin{figure}[htb]
	\centering
		\includegraphics[width=0.45\textwidth]{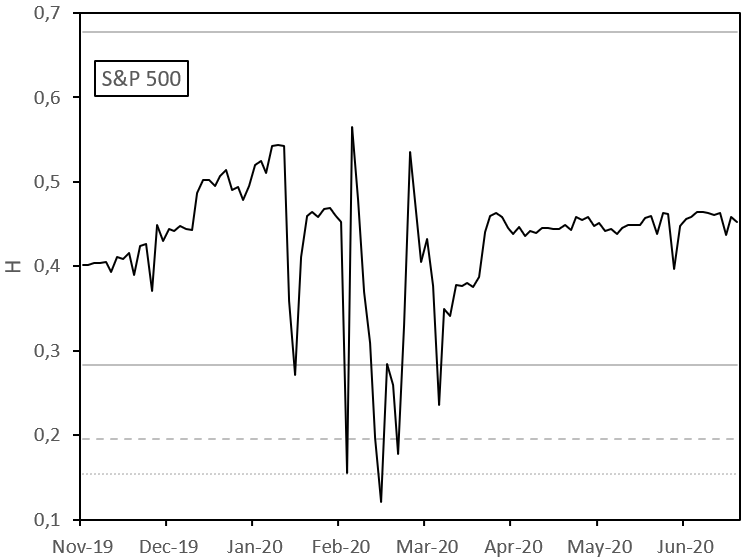} 
		\includegraphics[width=0.45\textwidth]{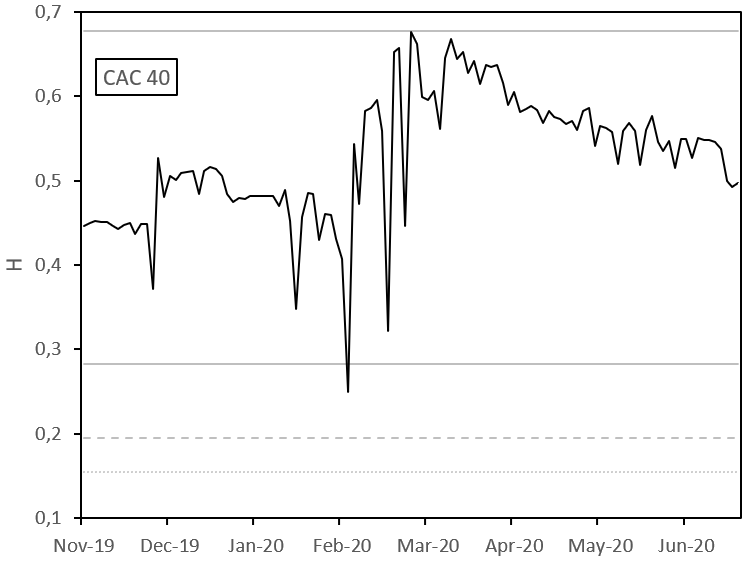} \\
		\includegraphics[width=0.45\textwidth]{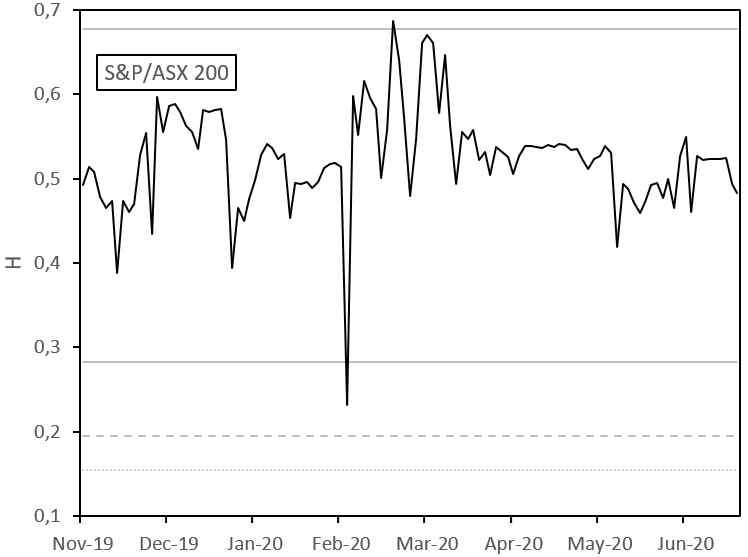} 
		\includegraphics[width=0.45\textwidth]{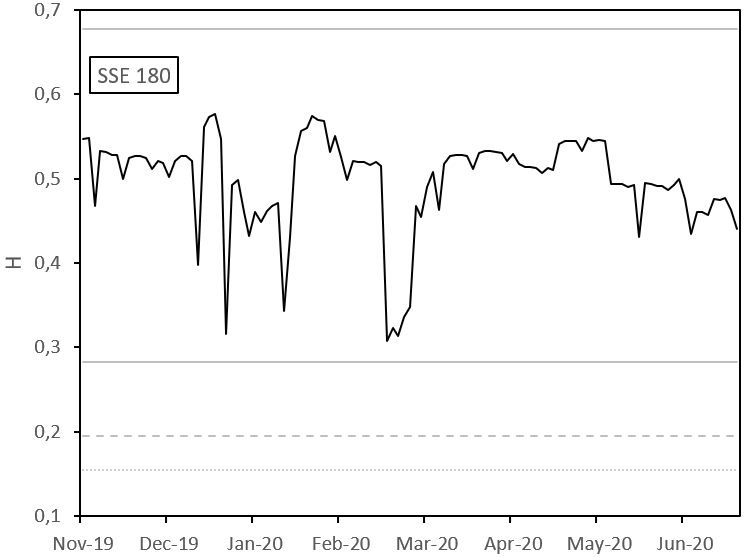} 
\begin{minipage}{0.7\textwidth}\caption{Daily evolution through time of the estimated Hurst exponent for S\&P 500 (top left), CAC 40 (top right), S\&P/ASX 200 (bottom left), and SSE 180 (bottom right) indices. The grey lines are the bounds of simulated confidence intervals with a confidence level of $95\%$, $99\%$, and $99.5\%$.}
	\label{fig:H}
\end{minipage}
\end{figure}

We display the time-varying Hurst exponent in Figure~\ref{fig:H} for the four focal stock indices of our study. Before the crisis, which begins in February or March with respect to the region, the Hurst exponent is not significantly different from $1/2$, so that we cannot reject $H_0$ during this period. During the crisis, we observe very low Hurst exponents for the S\&P 500 index which make it possible to reject $H_0$ with a confidence of more than $99\%$. But for the three other indices, the drawdown of the Hurst exponent is much less significant, in particular for the SSE 180 index. According to this approach, the market becomes clearly inefficient if we consider the S\&P 500 index, whereas we cannot ascertain the inefficiency of the three other indices. It is also worth noting that after a peak downward, the Hurst exponent reaches abnormally high values for the French and the Australian indices. It suggests a persistence of the crisis: a short mean-reverting phenomenon followed by positively correlated price returns.

\begin{figure}[htb]
	\centering
		\includegraphics[width=0.45\textwidth]{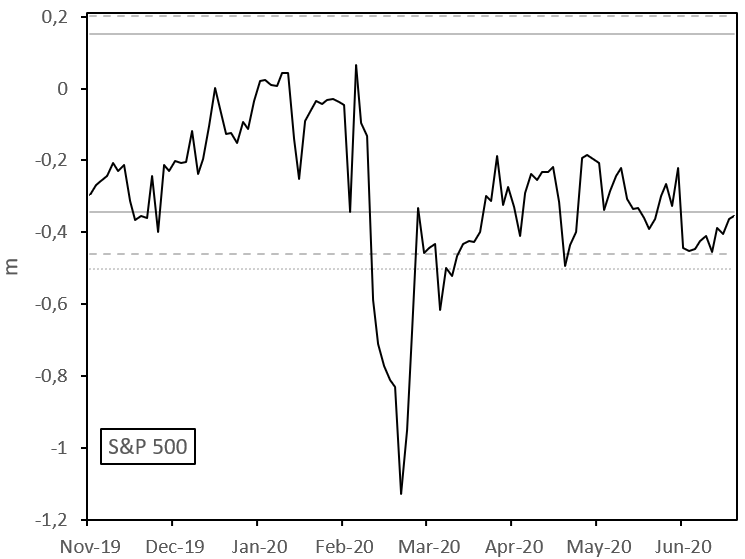} 
		\includegraphics[width=0.45\textwidth]{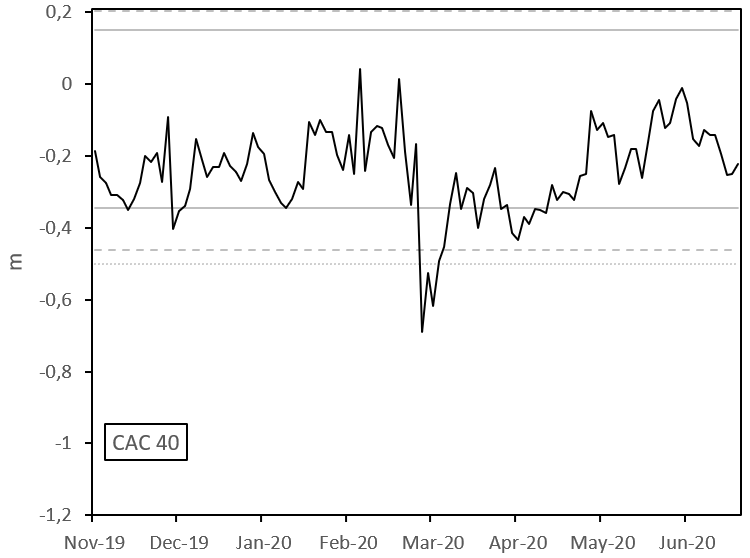} \\
		\includegraphics[width=0.45\textwidth]{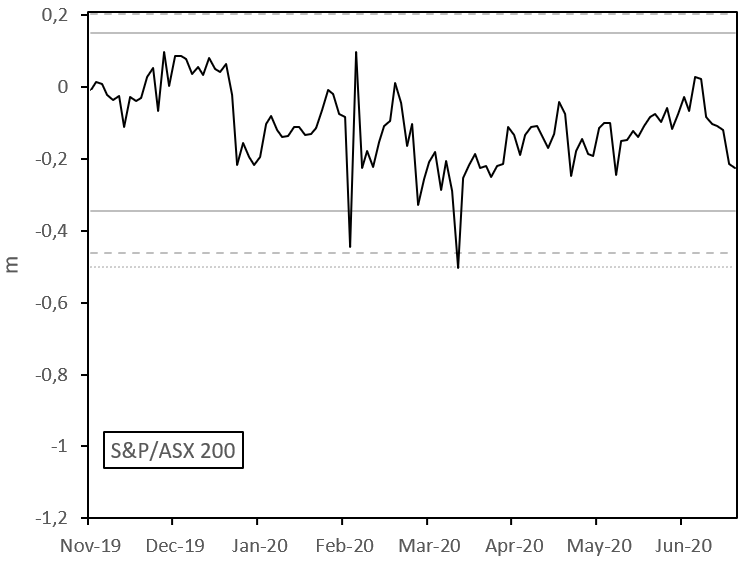} 
		\includegraphics[width=0.45\textwidth]{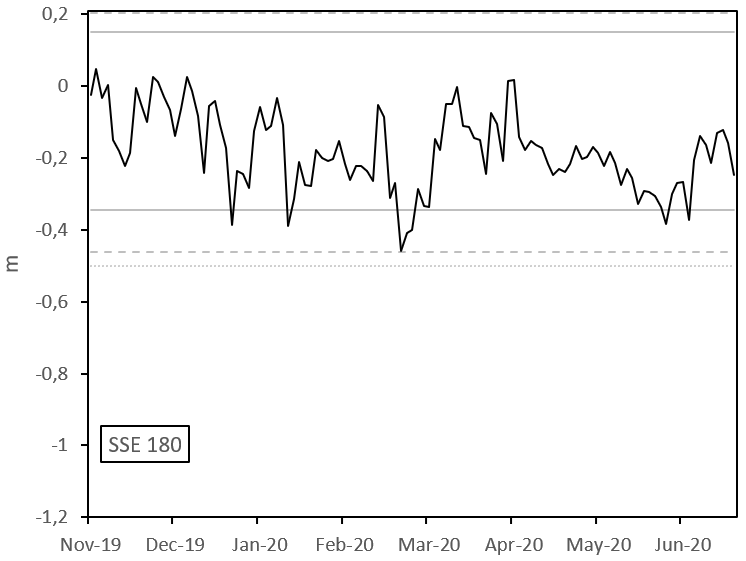} 
\begin{minipage}{0.7\textwidth}\caption{Daily evolution through time of the estimated $m$ parameter of an fLsm for S\&P 500 (top left), CAC 40 (top right), S\&P/ASX 200 (bottom left), and SSE 180 (bottom right) indices. The grey lines are the bounds of simulated confidence intervals with a confidence level of $95\%$, $99\%$, and $99.5\%$.}
	\label{fig:m}
\end{minipage}
\end{figure}

If we consider another indicator of market efficiency, namely the memory parameter $m$ of an fLsm, the conclusions regarding the impact of COVID-19 on the market efficiency are not the same. In Figure~\ref{fig:m}, we observe first that $m$ is in general negative, even before the crisis. It indicates a dominating mean-reversion phenomenon in the stock markets. But it is in fact often not significantly different from 0. When the crisis occurs, $m$ goes downward and $H_0$ is rejected with a confidence higher than $99\%$, whatever the stock index. The most significant drawdown of $m$ is again for the S\&P 500 index. The duration of the significant inefficiency varies among the indices. The longer period is for the S\&P 500 index. The S\&P/ASX 200 has a very short period of inefficiency.

\begin{figure}[htb]
	\centering
		\includegraphics[width=0.45\textwidth]{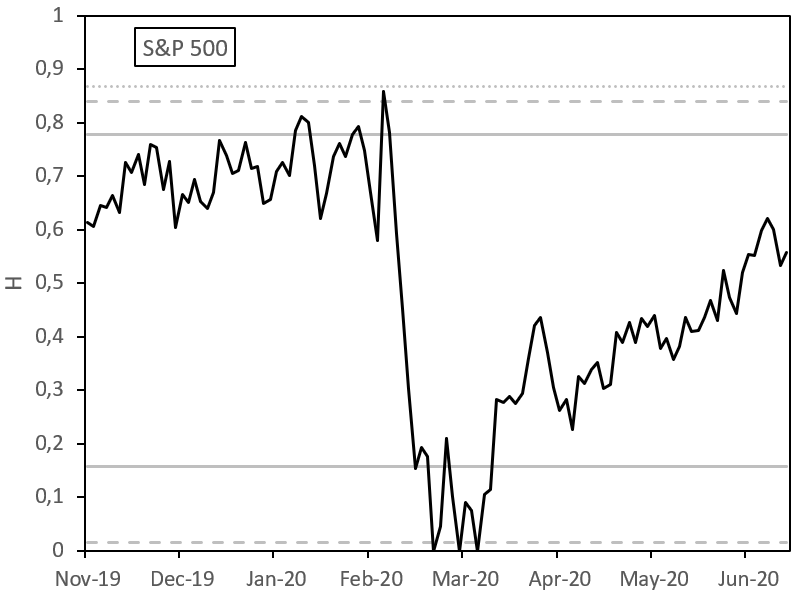} 
		\includegraphics[width=0.45\textwidth]{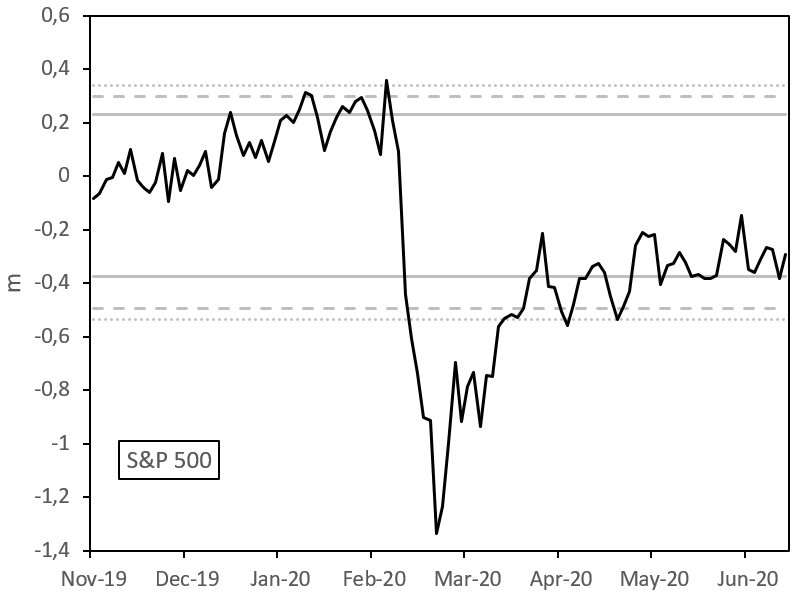} \\
		\includegraphics[width=0.45\textwidth]{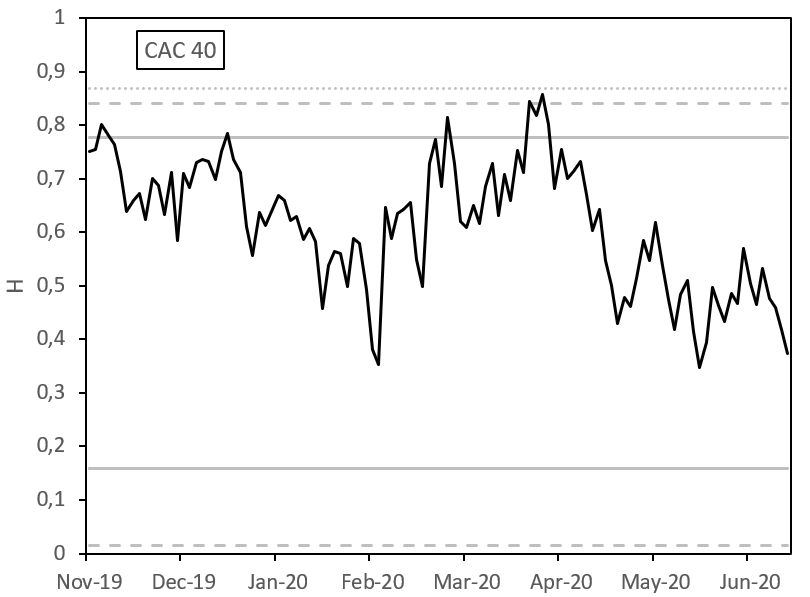} 
		\includegraphics[width=0.45\textwidth]{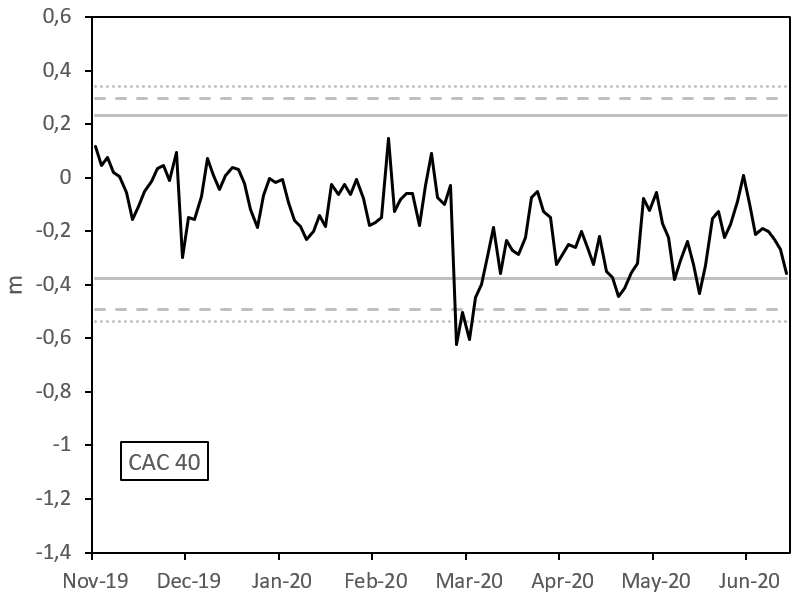} 
\begin{minipage}{0.7\textwidth}\caption{Daily evolution through time of the estimated $H$ (left) and $m$ (right) parameters for S\&P 500 (top) and CAC 40 (bottom) indices, with $k=0.5$ instead of 2. The grey lines are the bounds of simulated confidence intervals with a confidence level of $95\%$, $99\%$, and $99.5\%$.}
	\label{fig:k05}
\end{minipage}
\end{figure}

Using $k=0.5$ instead of $k=2$, as discussed in Section~\ref{sec:fLsm}, leads to similar results. In particular, the loss of efficiency is very strong for the S\&P 500 index, regardless of the indicator used, $H$ or $m$. For the CAC 40 index, the Hurst exponent is never significantly below $1/2$, but $m$ becomes significantly negative during the first lock-down. We display the corresponding evolutions in Figure~\ref{fig:k05}.

\begin{figure}[htb]
	\centering
		\includegraphics[width=0.45\textwidth]{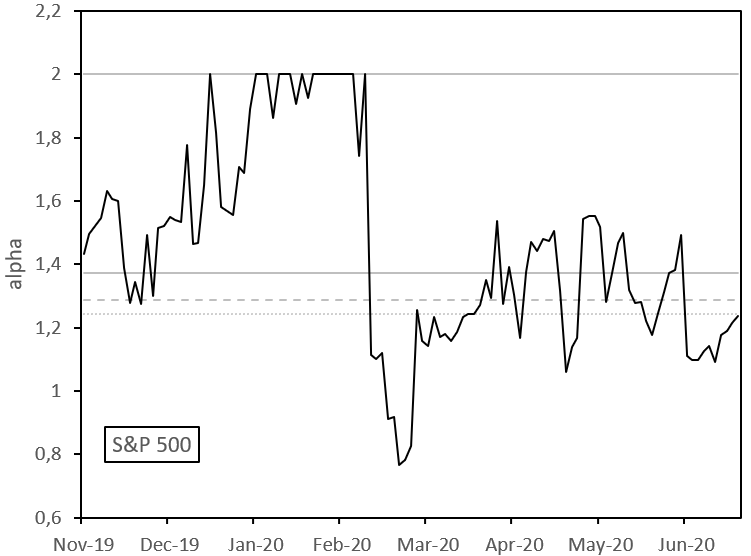} 
		\includegraphics[width=0.45\textwidth]{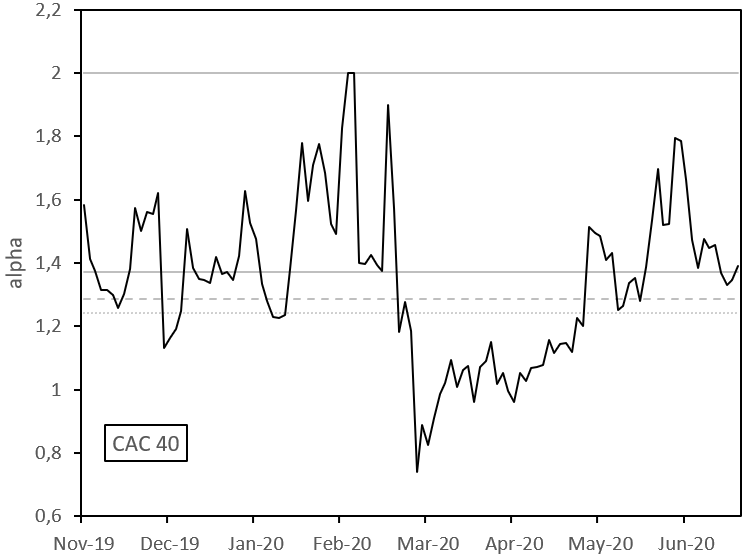} \\
		\includegraphics[width=0.45\textwidth]{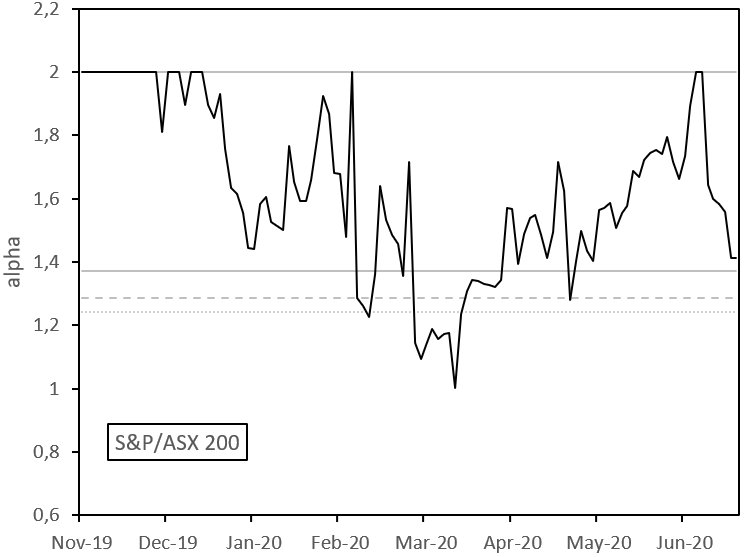} 
		\includegraphics[width=0.45\textwidth]{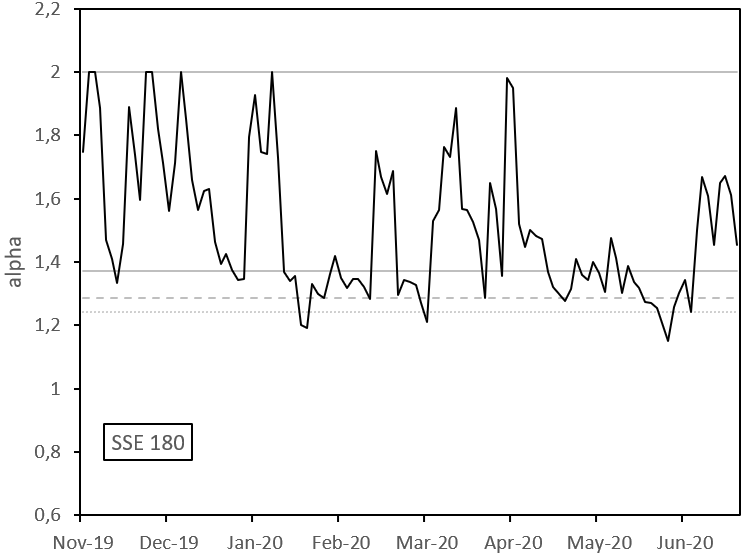} 
\begin{minipage}{0.7\textwidth}\caption{Daily evolution through time of the estimated $\alpha$ parameter of an fLsm for S\&P 500 (top left), CAC 40 (top right), S\&P/ASX 200 (bottom left), and SSE 180 (bottom right) indices. The grey lines are the bounds of simulated confidence intervals with a confidence level of $95\%$, $99\%$, and $99.5\%$.}
	\label{fig:alpha}
\end{minipage}
\end{figure}

It is also interesting to track another important parameter of the fLsm, namely $\alpha$, which depicts the size of the tails of the distribution of price returns. When $\alpha=2$, the price returns follow a Gaussian distribution. The lower $\alpha$, the fatter the tails. For the American, French, and Australian indices, we observe in Figure~\ref{fig:alpha} a negative impact of the crisis on $\alpha$. It means that extreme events tend to occur more frequently and with a larger magnitude. This stylized fact is confirmed by another approach relying on non-parametric densities~\cite{GKL}. The evolution of $\alpha$ in the Chinese market is not similar to the three other indices and the values reached are less significantly different from 2. A progressive recovery toward high values of $\alpha$ is visible for the CAC 40 and the S\&P/ASX 200 indices. For the S\&P 500 index, we also observe an abrupt increase of $\alpha$ after the peak of the crisis, but it is of limited amplitude and $\alpha$ remains at a fairly low value.

We display in Tables~\ref{tab:rangeH} and~\ref{tab:rangeM} the range of values reached during the period by the two efficiency indicators, $H$ and $m$, for the ten stock indices considered. These tables confirm that the greatest impact of COVID-19 on market efficiency occurred for US indices, whatever the efficiency indicator. We also note that inefficiency always leads to negative values of $m$. Indeed, the observed upper bounds in the period are never significantly different from 0. On the contrary, for the Hurst exponent, we find downward peaks below $1/2$ as well as upward peaks above $1/2$, so that it is unclear whether inefficiency leads to high or low values for $H$. In fact, the presence of fatter tails during the crisis biases this efficiency indicator. So, focusing on the $m$ indicator, we find fairly synchronized peaks for indices of the same region: on 10th March 2020 in the USA, on 16th March 2020 in Europe. Other regions seem less affected: the maximal $|m|$ is indeed lower in Asia and Australia. However, the biased $H$ indicator suggests a similar loss of efficiency in Europe and in Asia. This underpins again the relevance of the refinement of the efficiency indicator to take into account the kurtosis. The two efficiency indicators also lead to opposite conclusions when considering the situation at the end of the sample: according to $H$, markets are efficient again everywhere, whereas they are significantly inefficient in the USA and in Japan according to $m$.

\begin{table}[htbp]
\centering
\begin{tabular}{|l|cc|cc|c|}
\hline
Stock index & Min $H$ & Date of the min & Max $H$ & Date of the max & $H$ on $T$ \\
\hline
S\&P 500 & 0.122 & 2020-03-04 & 0.565 & 2020-02-25 & 0.452
  \\
S\&P 100 &  0.098 & 2020-03-04 & 0.565 & 2020-02-25 & 0.427 \\
\hline
EURO STOXX 50 & 0.288 & 2020-02-20 & 0.665 & 2020-03-12 & 0.488 \\
Euronext 100 & 0.269 & 2020-02-20 & 0.680 & 2020-03-12 & 0.491  \\
DAX &  0.296 & 2020-02-20 & 0.678 & 2020-03-12 & 0.502  \\
CAC 40 & 0.250 & 2020-02-20 & 0.676 & 2020-03-12 & 0.497  \\
\hline 
Nikkei 225 & 0.237 & 2020-01-08 & 0.677 & 2020-03-24 & 0.458  \\
KOSPI & 0.343 & 2020-01-23 & 0.757 & 2020-03-25 & 0.506  \\
SSE 180 & 0.307 & 2020-03-05 & 0.576 & 2019-12-18 & 0.440  \\
\hline
S\&P/ASX 200 & 0.231 & 2020-02-20 & 0.686 & 2020-03-09 & 0.482  \\
\hline
\end{tabular}
\begin{minipage}{0.7\textwidth}\caption{Range of values reached by the efficiency indicator $H$ between November 2019 and June 2020 for ten stock indices. $T$ is the 29th June 2020. An efficient market corresponds to $H$ close to $1/2$.}
\label{tab:rangeH}
\end{minipage}
\end{table}

\begin{table}[htbp]
\centering
\begin{tabular}{|l|cc|cc|c|}
\hline
Stock index & Min $m$ & Date of the min & Max $m$ & Date of the max & $m$ on $T$ \\
\hline
S\&P 500 & -1.128 & 2020-03-10 & 0.065 & 2020-02-25 & -0.355
  \\
S\&P 100 &  -1.065 & 2020-03-10 & 0.065 & 2020-02-25 & -0.464 \\
\hline
EURO STOXX 50 & -0.474 & 2020-03-18 & 0.043 & 2020-02-25 & -0.188 \\
Euronext 100 & -0.603 & 2020-03-16 & 0.055 & 2020-05-15 & -0.190  \\
DAX &  -0.576 & 2020-03-16 & 0.152 & 2020-03-09 & -0.250  \\
CAC 40 & -0.690 & 2020-03-16 & 0.043 & 2020-02-25 & -0.222  \\
\hline 
Nikkei 225 & -0.417 & 2020-06-29 & 0.139 & 2020-03-19 & -0.417  \\
KOSPI & -0.355 & 2020-06-24 & 0.148 & 2020-03-24 & -0.290  \\
SSE 180 & -0.460 & 2020-03-10 & 0.048 & 2019-11-06 & -0.247  \\
\hline
S\&P/ASX 200 & -0.504 & 2020-03-26 & 0.097 & 2020-02-25 & -0.225  \\
\hline
\end{tabular}
\begin{minipage}{0.7\textwidth}\caption{Range of values reached by the efficiency indicator $m$ between November 2019 and June 2020 for ten stock indices. $T$ is the 29th June 2020. An efficient market corresponds to $m$ close to $0$.}
\label{tab:rangeM}
\end{minipage}
\end{table}

\section{Conclusion}

We have shown to which extent the stock markets become inefficient during the COVID-19 crisis. The efficiency is clearly rejected in the case of the S\&P index. On the contrary, the Hurst exponent does not make it possible to conclude about a loss of efficiency for the CAC 40, SSE 180, and S\&P/ASX 200 indices. However, we have highlighted a limitation of this indicator: it does not take into account the dynamic kurtosis of price returns. Therefore, if we use the more appropriate memory parameter of an fLsm, we observe the occurrence of an inefficiency period almost at the beginning of the crisis, even though it is less noticeable for the Chinese and the Australian indices. 

We have also introduced in this paper the tools used for this analysis, namely the estimation of a dynamic stable distribution along with the estimation of the dynamic Hurst exponent and memory parameter of an fLsm. An important free parameter in this approach is the discount factor which is related to the speed at which the weight of past information decreases. We have used a selection rule based on the minimization of a criterion depicting the uniformity and the independence of the PITs, consistently with the literature on the validation of density forecasts.

\section*{Acknowledgement}

The authors deeply thank Akin Arslan, Thomas Barrat, and Sarah Bouabdallah for their valuable help in the implementation of some of the methods described in this paper. MG thanks the participants of the Econphysics Colloquium 2021 in Lyon for useful comments. 



\bibliographystyle{plain}
\bibliography{biblioDynEff}

\end{document}